\documentclass[10pt,twocolumn,showpacs,preprintnumbers,amssymb,pra]{revtex4}        
\usepackage{graphicx}
\usepackage{amsmath}
\usepackage{color}
\usepackage{dcolumn}
\usepackage{bm}
\usepackage{longtable}
\begin{document}

\title{Improving the fidelity of teleportation through noisy channels using weak measurement}

\author{T. Pramanik\footnote{tanu.pram99@bose.res.in}}
\affiliation{S. N. Bose National Centre for Basic Sciences, Salt Lake, Kolkata 700 098, India}

\author{A. S. Majumdar\footnote{archan@bose.res.in}}
\affiliation{S. N. Bose National Centre for Basic Sciences, Salt Lake, Kolkata 700 098, India}

\begin{abstract}
We employ the technique of weak measurement in order to enable preservation
of teleportation fidelity for two-qubit noisy channels. We consider
one or both qubits of a maximally entangled state to undergo amplitude
damping, and show that the application of weak measurement and a
subsequent reverse operation could lead to a fidelity greater than $2/3$ 
for any value of the decoherence parameter.  The success probability of
the protocol decreases with the strength of weak measurement, and is lower
when both the qubits are affected by decoherence. Finally, 
our protocol is shown to work for the Werner state too. 

\end{abstract}


\maketitle

\section{Introduction}

The primary goal of quantum information processing is to enable 
performing tasks that are unable to be accomplished classically.
Teleportation \cite{Tele1} is a typical information processing task
where at present there is intense activity in extending the
experimental frontiers \cite{TeleE1}. At the practical level teleportation
is implemented through the sharing of quantum entanglement by separated
parties involving the transmission of quantum particles over large
distances. Environmental interaction is a ubiquitous process here, which
unless controlled through well-devised means, leads to an inevitable loss
of fidelity of the teleported quantum states. Depending upon the magnitude
of environmental effects, the fidelity could fall below the maximum limit
attainable using classical means, thereby nullifying the quantum advantage of 
teleportation.

Though decoherence is generally responsible for the decay of quantum 
correlations in entangled states, and the associated loss of fidelity for
the corresponding information processing tasks for which such states are
utilized as resources, it has been noted that under certain specially chosen
conditions, it could also have a reverse effect. Entanglement between two
systems could be created or increased by their collective
interactions with a common environment \cite{bose}. Applications
of such effects in entanglement generation using trapped ions and
cavity fields have been suggested \cite{applic,biplab}. For the specific
case of teleportation it has been observed that the effect of amplitude
damping on one of the qubits of a shared bipartite state could lead to the
enhancement of fidelity above the classical limit  for a class of states 
whose fidelity lies just below quantum region \cite{Horo1}. However, such an
improvement is possible only for low values of the damping parameter, and
occurs only for a restricted class of states \cite{Som}.  

The preservation of entanglement in open systems is an important concern,
and in the present work we will approach this issue 
from another suggested direction. Recently, the
application of weak measurements has been suggested as a practically 
implementable method for protecting the fidelity of quantum states undergoing
decoherence through the amplitude damping 
channel \cite{WeakM1,WeakM2,WeakM3,WeakM4,WeakM5,WeakM6}. The original concept
of weak measurements proposed several years ago \cite{aav} showed
how it would be possible to get an experimental outcome outside the 
eigenvalue spectrum of an observable, if a sufficiently weak coupling of
the system and the apparatus along with the technique of post-selection is
employed. The idea of weak measurements has more recently been utilized in
several interesting applications 
such as demonstration of wave particle duality using cavity-QED experiments \cite{Wiseman_WPD},  superluminal propagation of light \cite{Brunner_2004},
observations of spin Hall effect
\cite{kwiat}, trajectories of photons \cite{stein}, direct measurement
of the quantum wave function \cite{lundeen}, and measurement of ultrasmall
time delays of light \cite{strubi}. 

The motivation of this work is to show how the fidelity of teleportation
using the resource of two qubits open to amplitude damping environments
could be protected with the help of weak measurement. For this purpose
we utilize the technique of weak measurement and its reversal as employed
recently in order to exhibit the suppression of the effect
of amplitude damping decoherence in preserving the entanglement of
two-qubit states \cite{WeakM1,WeakM2,WeakM3,WeakM4,WeakM5,WeakM6}. We first
study maximally entangled two-qubit channels which are the most
widely used resources in teleportation, and the effects of amplitude
damping on which have been investigated earlier for the purpose of 
obtaining fidelity greater than $2/3$  without using weak measurement \cite{Horo1,Som}. A fidelity below the classical limit of $2/3$ can be obtained with 
the help of shared randomness. It may be noted that with the help of
classical resources it is never possible to exceed the limit of $2/3$
 by employing any possible strategy  including post-selection \cite{classlim}.
We adapt the technique of weak measurement and its reversal in the
context of such a setting through the calculation of the optimal strength
of weak measurements required in order to maximize the output fidelity.
Finally, in order to show that the protocol of improving teleportation
fidelity using weak measurement and
its reversal is not restricted to the case of pure states, we also 
present an example of the mixed Werner state.

The plan of this paper is as follows. In Section II we provide a brief 
review of teleportation through two-qubit amplitude damping channels.
In Section III we present our analysis of employing weak measurements
in teleportation. 
As in any protocol involving weak measurement \cite{WeakM1,WeakM2,WeakM3,WeakM4,WeakM5,WeakM6,aav,kwiat,stein,lundeen,strubi}, the role of post-selection
is important here, since it allows one to work with only a sub-ensemble
of all the initial states.
Our analysis based on such post-selection shows that an average fidelity 
greater than $2/3$ is attainable for any strength of
decoherence for all maximally entangled two qubits if one of them undergoes
damping, while such a result holds for a sub-class of maximally entangled
two qubits if both of them are affected by decoherence. In Section IV
the Werner state is considered when both qubits interact with the environment,
and it is shown the teleportation fidelity may be enhanced to the quantum
region for a large range of the mixing parameter and the strength of 
decoherence.
We make some concluding remarks in Section V.     

\section{Teleportation through two-qubit amplitude damping channels}

In quantum teleportation with the help of entanglement the sender (say, Alice) 
is able to transfer the unknown quantum state of a qubit to a receiver 
(say, Bob), stationed at a distant location by performing local quantum 
operations and communicating two bits of classical information to Bob. The 
efficiency of teleportation, i.e, closeness of the teleported state with the 
initial state, $|\psi_i\rangle$ is determined by the fidelity $F$ 
 given by \cite{Fid1}
\begin{eqnarray}
F=\overline{\langle\psi_i|\sigma(|\psi_f\rangle)|\psi_i\rangle},
\end{eqnarray}
where $\sigma(\psi_f)$ is the density of the teleported state $|\psi_f\rangle$, 
and the average is taken over all initial states. For a given two-qubit 
entangled state $\sigma_{12}$ shared between Alice (who possesses the qubit 
labeled  `1') and Bob (who possesses the qubit labeled  `2'), the relation 
of the teleportation fidelity with the fully entangled fraction (FEF), 
$f(\sigma_{12})$ of $\sigma_{12}$ is given by\cite{SFrac}
\begin{eqnarray}
F(\sigma_{12})= \frac{2 f(\sigma_{12})+1}{3},
\end{eqnarray}
where $f(\sigma_{12})$ is defined by \cite{FSF}
\begin{eqnarray}
f(\sigma_{12})=\max_{|\phi\rangle} \langle\phi|\sigma_{12}|\phi\rangle,
\end{eqnarray}
with the maximization  taken over all two-qubit maximally entangled 
states $|\phi\rangle$. For the shared maximally entangled states $\sigma_{12}^M$, $f(\sigma_{12}^M)=1$ and $F(\sigma_{12}^M)=1$. In absence of entanglement, 
i.e., by using  shared randomness, the average teleportation fidelity achieved 
is $2/3$ \cite{CFid}. 

Let us suppose that Alice prepares two qubits in one of the four maximally 
entangled states, given by
\begin{eqnarray}
|\psi_{\pm}^M\rangle &=& \frac{|00\rangle_{12}\pm |11\rangle_{12}}{\sqrt{2}} \label{Parallel}  \\
|\phi_{\pm}^M\rangle &=& \frac{|01\rangle_{12}\pm |10\rangle_{12}}{\sqrt{2}} \label{anti-Parallel},
\end{eqnarray}
where subscript $i\in\{1,2\}$ represents the $i$-th qubit, and sends the second 
qubit to Bob. At the time of transit over the environment, the second qubit 
interacts with the environment. Due to this interaction, the entanglement between 
the qubits decreases and the maximally entangled state becomes a mixed 
state $\sigma_{12}$. If the FEF $f(\sigma_{12})\leq 1/2$, the state $\sigma_{12}$
is useless for the teleportation, as  one can achieve the fidelity $2/3$ on 
average classically.  

In Ref.\cite{Horo1}, the authors investigated whether 
using trace preserving LOCC (local operations and classical communications), 
one could get the quantum advantage, i.e., the fidelity to lie between $2/3$ 
and $1$ from the shared entangled state $\sigma_{12}$ with 
$f(\sigma_{12})\leq 1/2$. Any bistochastic map ($\Lambda$) which 
preserves both the trace and identity ($I$), i.e., ($\Lambda(I)=I$) fails to 
improve the FEF from the classical region ($0\leq f\leq 1/2$) to the quantum 
region ($f > 1/2$). Badziag et al.\cite{Horo1} showed that for a class of 
states $\rho_{12}$ given by 
 \begin{eqnarray}
\rho_{12}=
\begin{pmatrix}
\lambda_{11} & 0 & 0 & \lambda_{14}\\
0 & \lambda_{22} & -\gamma_{23} & 0\\
0 & -\gamma_{23} & \lambda_{33} & 0\\
\lambda_{14} & 0 & 0 & \lambda_{44} 
\end{pmatrix},
\label{Class_H}
\end{eqnarray}
where $\gamma_{23}\geq0$ and $\lambda_{14}$ are real; and $\lambda_{22}+\lambda_{33}\geq \frac{1}{2}$, $\gamma_{23}\geq (1-\lambda_{22}-\lambda_{33})/2$,  the fidelity ($F(\rho_{12}) =(1+\lambda_{22}+\lambda_{33}+2 \gamma_{23})/3\geq 2/3$) can be enhanced by applying 
a non-bistochastic map $\overline{\Lambda}$. For the choice of parameters $\lambda_{11}=\lambda_{14}=0$, $\lambda_{22}=3-2\sqrt{2}$, $\lambda_{33}=1$, $\lambda_{44}=2\sqrt{2}-2$ and $\gamma_{23}=\sqrt{2}-1$, the fidelity of the above state ($F(\rho_{12}) =2/3$, which belongs to classical region) can be enhanced up to $\frac{2.06}{3}$ (which lies in the quantum region) by applying $\overline{\Lambda}$ on any one 
of the qubits \cite{Horo1}.
The map $\overline{\Lambda}$ which represents the dissipative interaction of
any one qubit with the environment via the amplitude damping channel (ADC), is 
given by
\begin{eqnarray}
 \overline{\Lambda(\rho_{\alpha})} =
		 W_{\alpha,0}\rho_{\alpha} W_{\alpha,0}^{\dagger}
			 +W_{\alpha,1}\rho_{\alpha} W_{\alpha,1}^{\dagger},
\label{ADCMap}
\end{eqnarray}
where $\alpha\in\{1,2\}$, $\rho_{1(2)}=Tr_{2(1)}[\rho_{12}]$, and the operators $W_{\alpha,i}$ are given by
\begin{eqnarray}
W_{\alpha,0}= \begin{pmatrix}
1 & 0 \\ 
0 & \sqrt{\overline{D}_{\alpha}}
\end{pmatrix},       \hspace{1cm}
W_{\alpha,1}= \begin{pmatrix}
0 & \sqrt{D_\alpha} \\ 
0 & 0
\end{pmatrix} , 
\label{Deco1}
\end{eqnarray}
where $\overline{D}_{\alpha}=1-D_\alpha$. Here $D_1$ and $D_2$ are the strength of interactions of the 1st qubit (belonging to Alice) and the 2nd qubit (belonging to Bob) with the environment, respectively, and $\displaystyle\sum_i W_{\alpha,i}^{\dagger} W_{\alpha,i} =I$. The above map describes the interaction of the environment (which is initially in the state $|0\rangle_{E}$) with the qubit by the following transitions
\begin{eqnarray}
|0\rangle_i |0\rangle_E &\rightarrow & |0\rangle_i|0\rangle_E, \nonumber \\
|1\rangle_i |0\rangle_E &\rightarrow & \sqrt{\overline{D}_\alpha} |1\rangle_i|0\rangle_E + \sqrt{D_\alpha} |0\rangle_i|1\rangle_E,
\label{ADC}
\end{eqnarray}
where $i\in\{1,2\}$ and $\alpha=1 (2)$ for $i=1(2)$.

Later, in Ref.\cite{Som}, it was shown that the above interesting class of 
states $\rho_{12}$ (used in Ref.\cite{Horo1}) are obtained when Alice  
prepares the two-qubit maximally entangled state only in the class given by 
Eq.(\ref{Parallel}) and sends one qubit (say, the 2nd qubit) to Bob over ADC.
Now, for the purpose of enhancing the fidelity, Alice allows her qubit ($1$st 
qubit) also to 
interact with the environment via ADC.  According to Bandyopadhyay \cite{Som}, 
the fidelity is increased from the classical region to the quantum region due 
to the enhancement of classical correlations by the application of ADC on the
$1$st qubit as LOCC by itself is unable to increase the entanglement.
This protocol is not effective if the prepared maximally entangled state 
is chosen from the class given by Eq.(\ref{anti-Parallel}).

\section{Application of weak measurement and measurement reversal}

In earlier studies \cite{WeakM1,WeakM2,WeakM3,WeakM4,WeakM5} it has been shown 
that the effect of amplitude damping decoherence (given by Eq.(\ref{ADC})) can 
be suppressed by  weak quantum measurement and reversing quantum measurement 
(WMRQM) \cite{WeakM3}. In the present work we consider two cases. In the 1st case (``{\it Case I}"),  Alice prepares two qubits in one of the above two classes given by 
Eqs.(\ref{Parallel}) and (\ref{anti-Parallel}), and sends the 2nd qubit to Bob.
 Here the 2nd qubit is affected by ADC and 1st qubit is unaffected. In this 
case, for all shared states $\rho_{12}$ whose teleportation fidelities lie in 
the classical region, Alice and Bob are able to enhance the fidelity to the 
quantum region with the help of WMRQM, as we show below. In the second case 
(``{\it Case II}"), we consider both the 1st and 2nd particles to be interacting
 with environment. Here we show that for the class of states which are unable 
to achieve fidelity in the quantum region after allowing the interaction of 
Alice's particle with the environment,  the help of WMRQM enables attainment of 
fidelity above classical region. However, if the prepared state is chosen from the class 
given by Eq.(\ref{anti-Parallel}), the WMRQM technique fails to shift the 
fidelity from the classical region to the quantum region. We also calculate the 
the success probability (which is a consequence of the  non-unitary operation 
for the 
weak measurement) \cite{WeakM4} and show how it decreases with increment of 
the strength of the weak measurement.

Our protocol for both the cases proceeds as follows.
First, Alice prepares two qubits in one of the maximally entangled states 
given by Eqs.(\ref{Parallel}) and (\ref{anti-Parallel}). Before  allowing the 
interaction with environment via ADC, Alice makes a weak measurement with the 
strength $p_i$ on the $i$-th particle ($i=1,2$). The weak measurement is 
achieved by reducing the sensitivity of the detector, i.e., the detector 
clicks with probability $p_i$ if the input qubit is in the state $|1\rangle_i$, 
and never clicks if the input qubit is in the state $|0\rangle_i$ \cite{WeakM3,WeakM4}. When the detector clicks, the protocol fails as the input state 
collapses on the state $|1\rangle_i$ in an irreversible way. The success
 probability plays an important role in our protocol. When the detector does
 not click, the input state partially collapses  towards the state 
$|0\rangle_i$ which is unaffected by the interaction given by 
Eq.(\ref{ADC}) \cite{WeakM3}. The measurement operator corresponding to the 
detection of the particle is given by
\begin{eqnarray}
M_{\alpha,1}=\begin{pmatrix}
0 & 0\\
0 &\sqrt{p_{\alpha}}
\end{pmatrix},
\label{WKM1}
\end{eqnarray}
which does not have any inverse and hence, $M_{\alpha,1}$ is irreversible. The 
measurement operator that describes the situation when  the detector has not 
clicked is given by
\begin{eqnarray}
M_{\alpha,0}=\begin{pmatrix}
1 & 0\\
0 & \sqrt{\overline{p}_\alpha},
\end{pmatrix},
\label{WKM0}
\end{eqnarray}
where $\overline{p}_\alpha=1-p_\alpha$ and $M_{\alpha,0}^\dagger M_{\alpha,0} + M_{\alpha,1}^\dagger M_{\alpha,1}=I$. Here, $M_{\alpha,0}$ is the reversible having a 
mathematical inverse.

{\it Case I.} : Here, only the $2$nd qubit is affected by the amplitude damping
 decoherence when Alice sends it to Bob over the environment. To reduce the 
effect of ADC, Alice makes a weak measurement before sending the $2$nd qubit 
and after receiving it, Bob makes a reverse weak measurement. After making the 
weak measurement on the $2$nd qubit by Alice, the two-qubit state (unnormalized) 
becomes
\begin{eqnarray}
\rho^W_{\pm} &=& (I\otimes M_{2,0}) |\psi\rangle^M_\pm\langle\psi| (I\otimes M_{2,0}^{\dagger}) \nonumber \\
             &=& \frac{1}{2}\begin{pmatrix}
                 1 & 0 & 0 & \pm \sqrt{\overline{p}_2} \\
                 0 & 0 & 0 & 0 \\
                 0 & 0 & 0 & 0 \\
                 \pm \sqrt{\overline{p}_2} & 0 & 0 & \overline{p}_2
             \end{pmatrix}
\label{ParaWeak}
\end{eqnarray}
 and 
\begin{eqnarray}
\sigma^W_{\pm} &=& (I\otimes M_{2,0}) |\phi\rangle^M_\pm\langle\phi| (I\otimes M_{2,0}^{\dagger}) \nonumber \\
             &=& \frac{1}{2}\begin{pmatrix}
                 0 & 0 & 0 & 0 \\
                 0 & \overline{p}_2 & \pm \sqrt{\overline{p}_2} & 0 \\
                 0 & \pm \sqrt{\overline{p}_2} & 1 & 0 \\
                 0 & 0 & 0 & 0
             \end{pmatrix}
\label{AntiParaWeak}
\end{eqnarray}
when Alice prepares the initial state in the maximally entangled forms given 
by Eq.(\ref{Parallel}) and Eq.(\ref{anti-Parallel}), respectively. Here the 
detector's inefficiency, or the success probability is given by 
\begin{eqnarray}
P^D_2=Tr[\rho^W_{\pm}]=Tr[\sigma^W_{\pm}]= (1-\frac{p_2}{2}).
\label{PSuccD2}
\end{eqnarray}

Next, Alice sends the 2nd qubit to Bob over ADC. Due to the effect of the 
interaction, the shared state $\rho_\pm^W$ becomes 
\begin{eqnarray}
\rho_\pm^D &=& (I\otimes W_{2,0}) \rho_{\pm}^W (I\otimes W_{2,0}^{\dagger}) + (I\otimes W_{2,1}) \rho_{\pm}^W (I\otimes W_{2,1}^{\dagger}) \nonumber \\
&=& \frac{1}{2}\begin{pmatrix}
1 & 0 & 0 & k1 \\
0 & 0 & 0 & 0 \\
0 & 0 & D_2 \overline{p}_2 & 0 \\
k1 & 0 & 0 & k1^2
\end{pmatrix},
\label{ParaDeco}
\end{eqnarray}
where $k1=\pm \sqrt{\overline{D}_2\overline{p}_2}$. Similarly, $\sigma_\pm^W$ 
becomes
\begin{eqnarray}
\sigma_\pm^D &=& (I\otimes W_{2,0}) \sigma_{\pm}^W (I\otimes W_{2,0}^{\dagger}) + (I\otimes W_{2,1}) \sigma_{\pm}^W (I\otimes W_{2,1}^{\dagger}) \nonumber \\
&=& \frac{1}{2}\begin{pmatrix}
D_2 \overline{p}_2 & 0 & 0 & 0 \\
0 & k1^2 & k1 & 0 \\
0 & k1 & 1 & 0 \\
0 & 0 & 0 & 0
\end{pmatrix}.
\label{AntiDeco}
\end{eqnarray}
Finally, Bob applies the reverse quantum measurement \cite{WeakM3} $N_{2,0}$ 
(corresponding to $M_{2,0}$ given in Eq.(\ref{WKM0})) given by
\begin{eqnarray}
N_{2,0}=\begin{pmatrix}
\sqrt{\overline{q}_2} & 0\\
0 & 1 
\end{pmatrix},
\label{RMB0}
\end{eqnarray}
where $q_2$ is the strength of the weak measurement on the $2$nd qubit and 
$\overline{q}_2=1-q_2$. At the end, Alice and Bob  actually share one of the 
states given by 
\begin{eqnarray}
\rho_\pm^R &=& (I\otimes N_{2,0}) \rho_\pm^D (I\otimes N_{2,0}^{\dagger})  
\label{ParaFState1}\\
&=&                                 
\begin{pmatrix}
\frac{\overline{q}_2}{2} & 0 & 0 & \frac{\pm \sqrt{\overline{D}_2 \overline{p}_2 \overline{q}_2}}{2}  \\
0 & 0 & 0 & 0 \\
0 & 0 & \frac{D_2 \overline{p}_2 \overline{q}_2}{2} & 0 \\
\frac{\pm \sqrt{\overline{D}_2 \overline{p}_2 \overline{q}_2}}{2} & 0 & 0 & \frac{\overline{D}_2 \overline{p}_2}{2}
\end{pmatrix}\nonumber 
\\
\sigma_\pm^R  &=& (I\otimes N_{2,0}) \sigma_\pm^D (I\otimes N_{2,0}^{\dagger})  
\label{AntiParaFState1}\\
&=& 
\begin{pmatrix}
\frac{ D_2\overline{p}_2 \overline{q}_2}{2} & 0 & 0 & 0 \\
0 & \frac{\overline{D}_2 \overline{p}_2}{2} & \frac{\pm \sqrt{\overline{D}_2 \overline{p}_2 \overline{q}_2}}{2} & 0 \\
0 & \frac{\pm \sqrt{\overline{D}_2 \overline{p}_2 \overline{q}_2}}{2} & \frac{\overline{q}_2}{2} & 0 \\
0 & 0 & 0 & 0 
\end{pmatrix}. \nonumber
\end{eqnarray}

The FEFs are equal for both the states given by Eqs.(\ref{ParaFState1}) and 
(\ref{AntiParaFState1}), i.e., the FEF is the same whether Alice prepares 
the initial two-qubit state in the class given by Eq.(\ref{Parallel}) or 
Eq.(\ref{anti-Parallel}), and it is given by
\begin{eqnarray}
f_1 = \frac{\overline{p}_2+\overline{q}_2+2\sqrt{\overline{D}_2\overline{p}_2\overline{q}_2}-D_2\overline{p}_2}{2(\overline{p}_2+\overline{q}_2)-2 D_2 q_2\overline{p}_2}
\label{Non-OptimalFEF1}
\end{eqnarray}
The strength of the weak measurement has to be chosen so as to achieve the
purpose of the experiment. In Ref.\cite{WeakM6}  the authors calculate the
optimum strength of the weak measurement that maximizes the 
concurrence of the non-maximally entangled state used by them in order to 
protect the entanglement from the interaction of the qubits with environment 
via ADC. The optimal value of $q_2$ which maximally protects
 the fidelity of the unknown teleported state undergoing amplitude damping is obtained by 
 maximizing 
$ f_1$ (given by Eq.(\ref{Non-OptimalFEF1})) with respect to $q_2$.
It turns out that for both the classes of prepared states, the 
optimal strength, $q_2^O$ of the reverse measurement is the same, and is 
given by
\begin{eqnarray}
q_2^O= \frac{3 D_2 \overline{p}_2+D_2^2 \overline{p}_2^2+p_2}{(1+D_2 \overline{p}_2)^2}.
\label{Optimal1}
\end{eqnarray}
Note that though the choice of $q_2=p_2+D_2\overline{p_2}$ optimally preserves 
the entanglement of the maximally entangled state \cite{WeakM5,WeakM6}, it 
does not maximize the fidelity of the state passing through the noisy channel. 
For the choice of initial state from the class given by Eqs.(\ref{Parallel}) 
and (\ref{anti-Parallel}), 
 using Eqs.(\ref{Non-OptimalFEF1}) and (\ref{Optimal1}) one can calculate 
the optimal FEF 
which is given by
\begin{eqnarray}
f_1^O=\frac{2+D_2 \overline{p}_2}{2+2 D_2 \overline{p}_2},
\label{FEF1}
\end{eqnarray}
where $f_1^O$ is bounded by $0.75$ (occurs for the choice $D_2=1$ and $p_2=0$) 
and $1$ (occurs for either $p_2=1$, or $D_2=0$). Here one may note that the 
 optimal teleportation fidelity $F_1^O (=\frac{2 f_1^O + 1}{3})$ always 
belongs to 
the quantum
 region $(> 2/3)$ irrespective of the strength of decoherence. Due to the weak 
measurement and the reverse weak measurement, the overall success probability, 
i.e., the probability of obtaining the state $\rho_\pm^R$ ($\sigma_\pm^R$) when Alice prepares the two-qubit state in the class given by Eq.(\ref{Parallel}) (Eq.(\ref{anti-Parallel})) is given by \cite{WeakM4}
\begin{eqnarray}
P_{Succ}^1 &=& Tr[\rho_\pm^R]=Tr[\sigma_\pm^R] =\frac{1}{2} (\overline{p}_2+\overline{q}_2^O-D_2\overline{p}_2)  \nonumber \\
           &=&\frac{(1-D_2) (1-p_2) (2+D_2 (1-p_2))}{2+2D_2 (1-p_2)},
\label{PSucc1}
\end{eqnarray}
where the success probability lies  between $0$ (which occurs for either 
$D_2=1$, or $p_2=1$, or both) and $1$ (which occurs when both $D_2=0$ and $p_2=0$ hold simultaneously). 

Now, let us consider the situation when no weak measurement and its reverse is 
performed. Due to the effect 
of interaction on the $2$nd 
particle with the environment via ADC, the FEF of the two-qubit 
state prepared in one of the two classes of maximally entangled states given by 
Eqs.(\ref{Parallel}) and (\ref{anti-Parallel}), is given by\cite{Som}
\begin{eqnarray}
\overline{f}_1=\frac{1}{4} + \frac{1}{2} \sqrt{1-D_2}+\frac{1}{4} (1-D_2)
\label{FEFD1}
\end{eqnarray}
and the corresponding fidelity turns out to be $\overline{F}_1 =(2 \overline{f}_1+1)/3$. In the range $2\sqrt{2}-2\leq D_2 \leq 1$, the teleportation fidelity 
$\overline{F}_1$ lies in the classical region, and for others values, i.e., 
$0\leq D_2 < 2\sqrt{2}-2$, $\overline{F}_1$ lies in the quantum region.  In the  figure, FIG. 1 we compare the $F_1^O$ with the $\overline{F}_1$. One sees that
for sufficiently strong environmental interaction, the fidelity could
fall below the quantum region without the help of weak measurement. However,
as detailed in our protocol above, when one performs weak measurement 
and its subsequent reversal, the fidelity is preserved above the classical
value for all strengths of decoherence. This result holds irrespective of
whether the initial state is chosen to belong to the class given by 
Eq.(\ref{Parallel}) or by Eq.(\ref{anti-Parallel}).

It is interesting to note that the role of the reverse weak measurement done 
by Bob is more important than the weak measurement made by Alice before sending 
the $2$nd particle to Bob over the ADC. To see this point, we consider that 
Alice sends the $2$nd particle to Bob without making any weak measurement on it,
 i.e., $p_2=0$, over the environment. After getting the $2$nd particle, Bob makes
 an optimal weak measurement given by Eq.(\ref{RMB0}) with $q_2=q_2^O$ given 
by Eq.(\ref{Optimal1}). The optimal FEF in this case is given by
\begin{eqnarray}
f_{12}^O=\frac{2+D_2}{2+2D_2},
\end{eqnarray}
which is obtained from Eq.(\ref{FEF1}) by putting $\overline{p}_2=1$, and the 
 corresponding success probability is $\frac{2-D_2-D_2^2}{2(1+D_2)}$. 
Here, $F_{12}^O$ ($=(2 f_{12}^O+1)/3$) is not only greater than $\overline{f}_1$,
 but, also $F_{12}^O$ lies in the quantum region, i.e.,  $5/6 \leq F_{12}^O\leq 1$
 for all values of the decoherence parameter $D_2$ which lie in the region 
given by $1\geq D_2\geq 0$.

\begin{figure}[h]
{\rotatebox{0}{\resizebox{9.0cm}{5.0cm}{\includegraphics{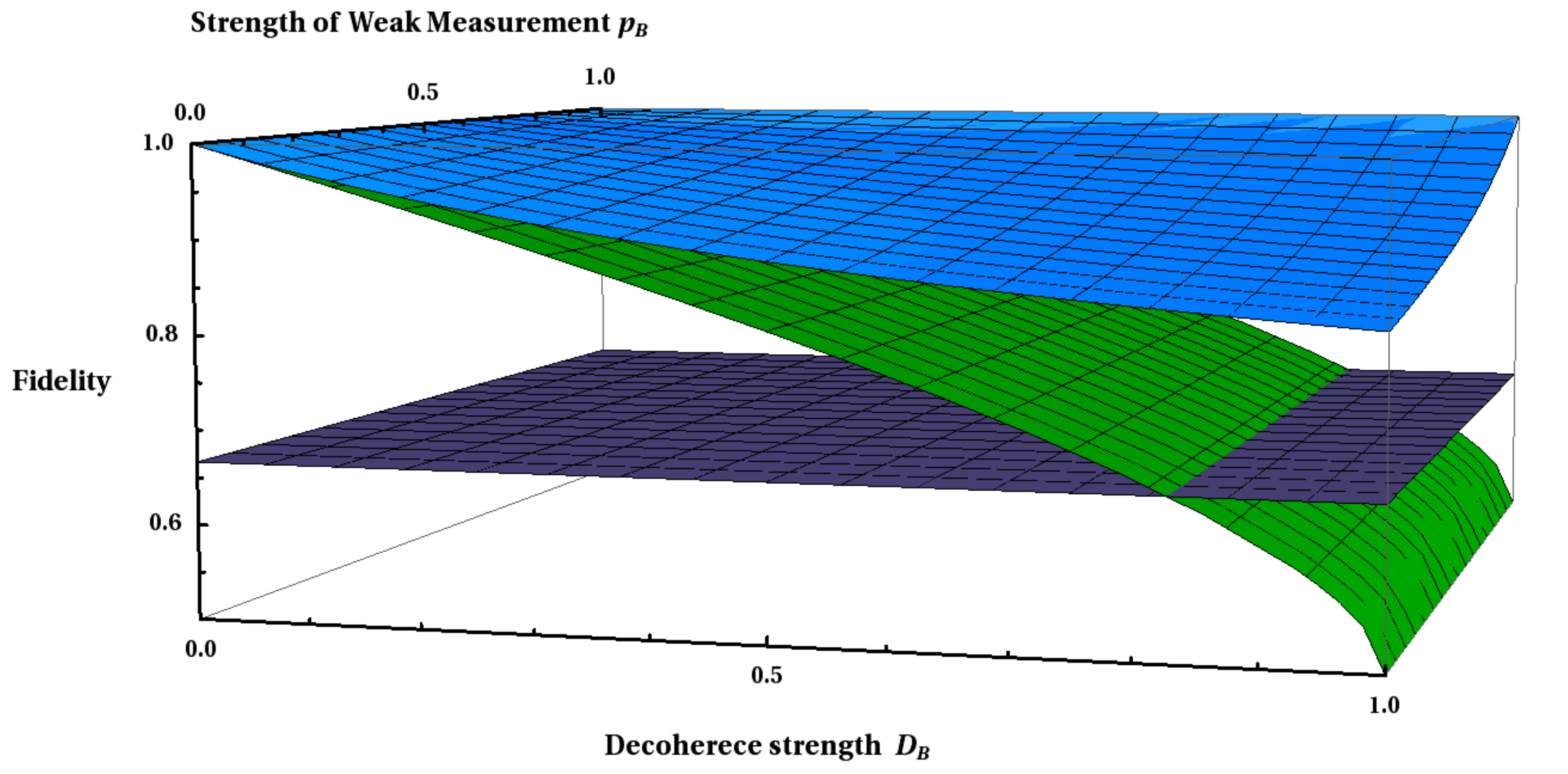}}}}

\caption{\footnotesize (Coloronline) The flat plane represents the average 
classical fidelity $\frac{2}{3}$. The surface intersecting it represents the 
fidelity $\overline{F}_1$ corresponding to the FEF $\overline{f}_1$ given by Eq.(\ref{FEFD1}). The uppermost surface 
represents 
the fidelity $F_1^O$ corresponding to the FEF $f_1^O$ given by Eq.(\ref{FEF1}).}
\end{figure}

{\it Case II. :} In this case, both the 1st and 2nd particles interact with 
the environment via ADC. To prevent the loss of information about the unknown 
state in the teleportation protocol, Alice makes weak measurements (given by 
Eq.(\ref{WKM0})), separately on each qubit. When the prepared two-qubit state 
belongs to the class given by Eq.(\ref{Parallel}), after the weak measurement the state becomes
\begin{eqnarray}
\rho_\pm^{WW} &=&  (M_{1,0}\otimes M_{2,0}) |\psi\rangle^M_\pm\langle\psi| (M_{1,0}^{\dagger} \otimes M_{2,0}^{\dagger}) \nonumber \\
&=& 
\begin{pmatrix}
\frac{1}{2} & 0 & 0 & \pm \frac{ \sqrt{\overline{p}_1 \overline{p}_2}}{2} \\
0 & 0 & 0 & 0 \\
0 & 0 & 0 & 0 \\
\pm \frac{ \sqrt{\overline{p}_1 \overline{p}_2}}{2} & 0 & 0 & \frac{\overline{p}_1 \overline{p}_2}{2}
\end{pmatrix}
\label{ParaWeak2}
\end{eqnarray}
Similarly when the state chosen is from the class given by Eq.(\ref{anti-Parallel}), after weak measurement the state becomes
\begin{eqnarray}
\sigma_\pm^{WW} =  
\begin{pmatrix}
0 &0 & 0 & 0 \\
0 & \frac{\overline{p}_2}{2}  & \pm\frac{ \sqrt{\overline{p}_1 \overline{p}_2}}{2}  & 0 \\
0 & \pm\frac{ \sqrt{\overline{p}_1 \overline{p}_2} }{2} & \frac{\overline{p}_1}{2}  & 0 \\
0 & 0 & 0 & 0 
\end{pmatrix}.
\label{ParaWeak2}
\end{eqnarray}
The corresponding success probabilities of the weak measurements are given by
\begin{eqnarray}
P_{12}^D(\rho_\pm^{WW})=Tr[\rho_\pm^{WW}]=\frac{1}{2}(1+\overline{p}_1 \overline{p}_2)
\label{PSuccD121}
\end{eqnarray}
and 
\begin{eqnarray}
P_{12}^D(\sigma_\pm^{WW})=Tr[\sigma_\pm^{WW}]=\frac{1}{2}(\overline{p}_1+ \overline{p}_2),
\label{PSuccD121}
\end{eqnarray}
respectively. 

Here, Alice sends the $2$nd qubit through the ADC and also allows her 
qubit (1st qubit) to interact with the environment. Hence, both particles 
interact with the environment via ADC. After the interaction with the 
environment, the noisy shared state takes one of the following forms (depending
upon the initial state)
\begin{eqnarray}
\rho_\pm^{DD} &=&\begin{footnotesize}                                               
\begin{pmatrix}
\frac{1+D_1 D_2 k4^2}{2} & 0 & 0 & \pm \frac{k5 k4}{2} \\            
0 & \frac{D_1 \overline{D}_2 k4^2}{2} & 0 & 0 \\
0 & 0 & \frac{\overline{D}_1 D_2 k4^2}{2} & 0 \\
\pm \frac{k5 k4}{2} & 0 & 0 &  \frac{k5^2 k4^2}{2}                            
\end{pmatrix}
\end{footnotesize} 
\label{ParaDeco2}\\
\sigma_\pm^{DD} &=&\begin{footnotesize}                                               
\begin{pmatrix}
\frac{D_1 \overline{p}_1}{2} + \frac{D_2 \overline{p}_2}{2} & 0 & 0 & 0 \\            
0 & \frac{\overline{D}_2 \overline{p}_2}{2} & \pm \frac{k4 k5}{2} & 0 \\
0 & \pm\frac{k4 k5}{2} & \frac{\overline{D}_1 \overline{p}_1}{2}  & 0 \\
0 & 0 & 0 & 0                            
\end{pmatrix}
\end{footnotesize}
\label{AntiParaDeco2},
\end{eqnarray}
where $k4 = \sqrt{\overline{p}_1 \overline{p}_2}$ and $k5=\sqrt{\overline{D}_1 \overline{D}_2}$. 
Next, both Alice and Bob apply reverse weak measurement with the 
strengths $q_1$ and $q_2$ on their respective particles.
Let us consider the two classes of initial states separately.

If the initial state is chosen to be  from the class given by 
Eq.(\ref{Parallel}),  the joint state now becomes
\begin{eqnarray}
\rho_\pm^{RR} &=& (N_{1,0} \otimes N_{2,0}) \rho_\pm^{DD} (N_{1,0}^{\dagger} \otimes N_{2,0}^{\dagger}) 
\label{ParaFState2}
\end{eqnarray}
where $N_{2,0}$ is given by Eq.(\ref{RMB0}) and Alice's reverse weak 
measurement operator $N_{1,0}$ is given by
\begin{eqnarray}
N_{1,0}=\begin{pmatrix}
\sqrt{\overline{q}_1} & 0\\
0 & 1 
\end{pmatrix}.
\label{RMA0}
\end{eqnarray}
Before maximizing the fidelity $f(\rho_{\pm}^{RR})$ in this case, for simplicity,
let us make the following assumptions. We consider $D_1=D_2=D$, i.e., both the 
qubits interact with similar environments, and also, $p_1=p_2=p$, i.e., the 
strength of weak measurements on both qubits are the same, and $q_1=q_2=q$,
as well. Similar to `{\it Case I}', we maximally enhance the teleportation 
fidelity (i.e., the FEF $f(\rho_{\pm}^{RR})$) by maximizing $f(\rho_{\pm}^{RR})$ 
with respect to the reverse weak measurement strength $q$. The optimal  FEF is 
given by 
\begin{eqnarray}
f_2^O=f(\rho_{\pm}^{RR})=\frac{1+\sqrt{1+D^2\overline{p}^2}+D^2\overline{p}^2}{2(1+D\overline{p}\sqrt{1+D^2\overline{p}^2}+D^2\overline{p}^2)},
\label{FEF2}
\end{eqnarray}
which occurs for the choice 
\begin{eqnarray}
q^O=\frac{1+D^2\overline{p}^2-\sqrt{\overline{D}^2\overline{p}^2(1+D^2\overline{p}^2)}}{1+D^2\overline{p}^2}.
\label{Optimal2}
\end{eqnarray}
From the above expression it follows that
 $f_2^O$ always lies in the quantum region, i.e.,  between $0.5$ (corresponding to $D=1$ and $p_2=0$) and $1.0$ (corresponding to $D=0$ and $p_2=0$). 
Simultaneously, the success probability decreases according to the 
relation
\begin{eqnarray}
P^{2}_{Succ}&=&Tr[\rho_{\pm}^{RR}]  
\label{PSucc21} \\
&=&\frac{1}{1+D^2(1-p)^2}((1-D)^2(1-p)^2 (1+ \nonumber \\
&&D(1-p)\sqrt{1+D^2 (1-p)^2}+D^2(1-p)^2)),\nonumber 
\end{eqnarray}
where we use  $q=q^O$. The success probability $P^{2}_{Succ}$ varies from $0$ to $1$.

 \begin{figure}[h]
{\rotatebox{0}{\resizebox{9.0cm}{5.0cm}{\includegraphics{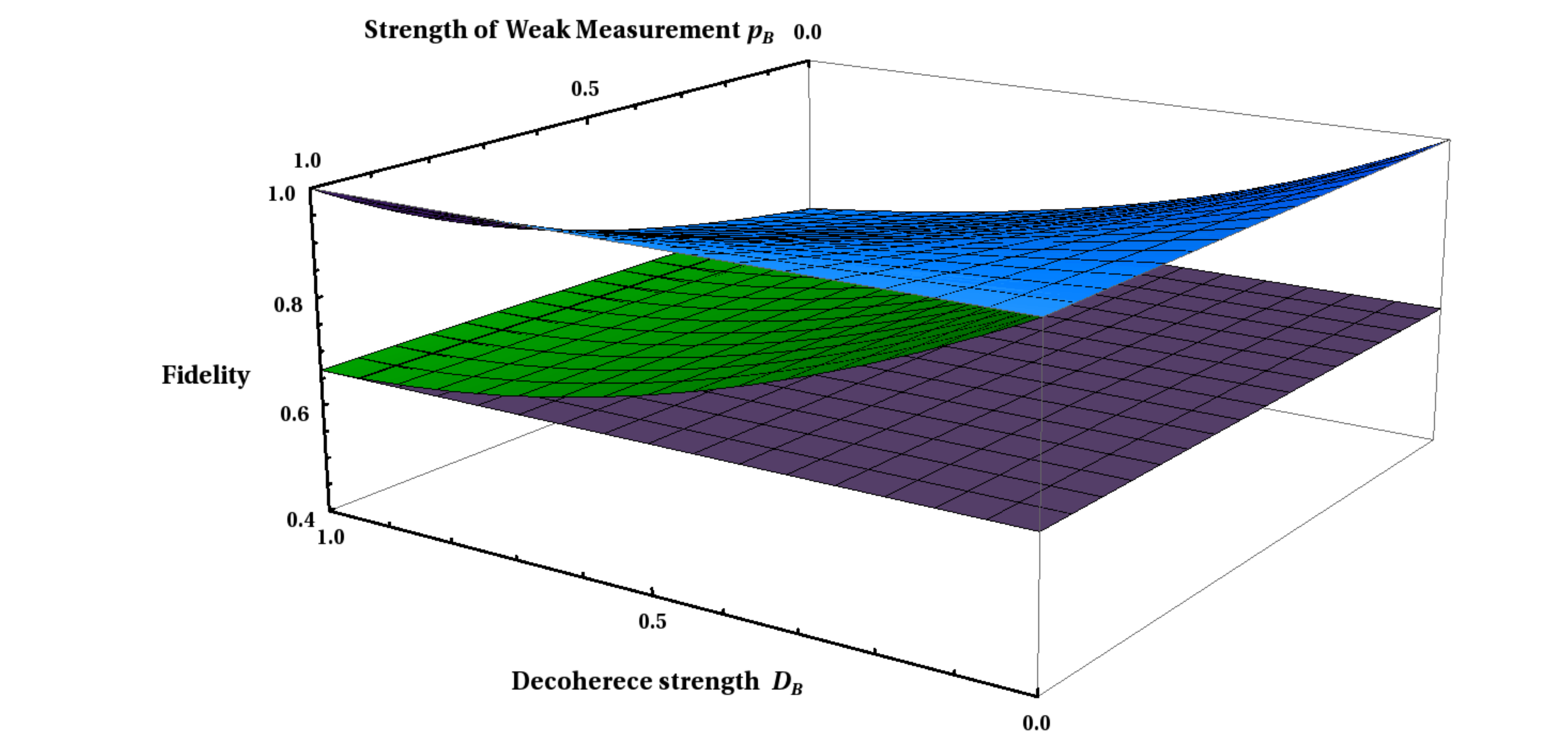}}}}

\caption{\footnotesize (Coloronline) The flat plane represents the average 
classical fidelity $\frac{2}{3}$. The lower surface represents the fidelity $\overline{F}_2$ corresponding to the FEF $\overline{f}_2$ given  by Eq.(\ref{FEFD2}). The 
upper surface represents the fidelity $F_2^O$ corresponding to the FEF $f_2^O$ given by Eq.(\ref{FEF2}).}
\end{figure}

Here again, we compare the above situation with the case when decoherence
acts without introducing weak measurement and reversal.
In the absence of weak measurement, when both the qubits undergo damping,
the FEF is given by \cite{Som}
\begin{eqnarray}
\overline{f}_2=1-D+\frac{D^2}{2}
\label{FEFD2}
\end{eqnarray}
and $\overline{F}_2$ is the corresponding fidelity. Note that when $D$ is 
chosen in
the range $2\sqrt{2}-2\leq D\leq 1$, though $\overline{F}_1$  lies in the 
classical region, it was shown that $\overline{F}_2$ is 
quantum \cite{Som,Horo1}. In FIG. 2, we compare the  optimal fidelity $F_2^O$ achieved 
using weak measurement and reversal with $\overline{F}_2$. The comparison 
shows that the weak measurement technique enhances the fidelity $F_2^O$ above 
$\overline{F}_2$ for the whole range of the decoherence parameter. 

Next, we compare the success probabilities for both the cases studied,
which are given by Eqs.(\ref{PSucc1}) and (\ref{PSucc21}), respectively. 
In FIG. 3,
 we plot the success probabilities $P^1_{Succ}$ with $P^2_{Succ}$ , as functions
of the decoherence parameter and the strength of weak measurement. Note
that in both the cases the corresponding success probabilities fall with
the increase of these parameter values. However, $P^1_{Succ}$ always lies
above $P^2_{Succ}$, since in the latter case both qubits undergo damping,
and two weak measurements are required.

\begin{figure}[h]
{\rotatebox{0}{\resizebox{9.0cm}{5.0cm}{\includegraphics{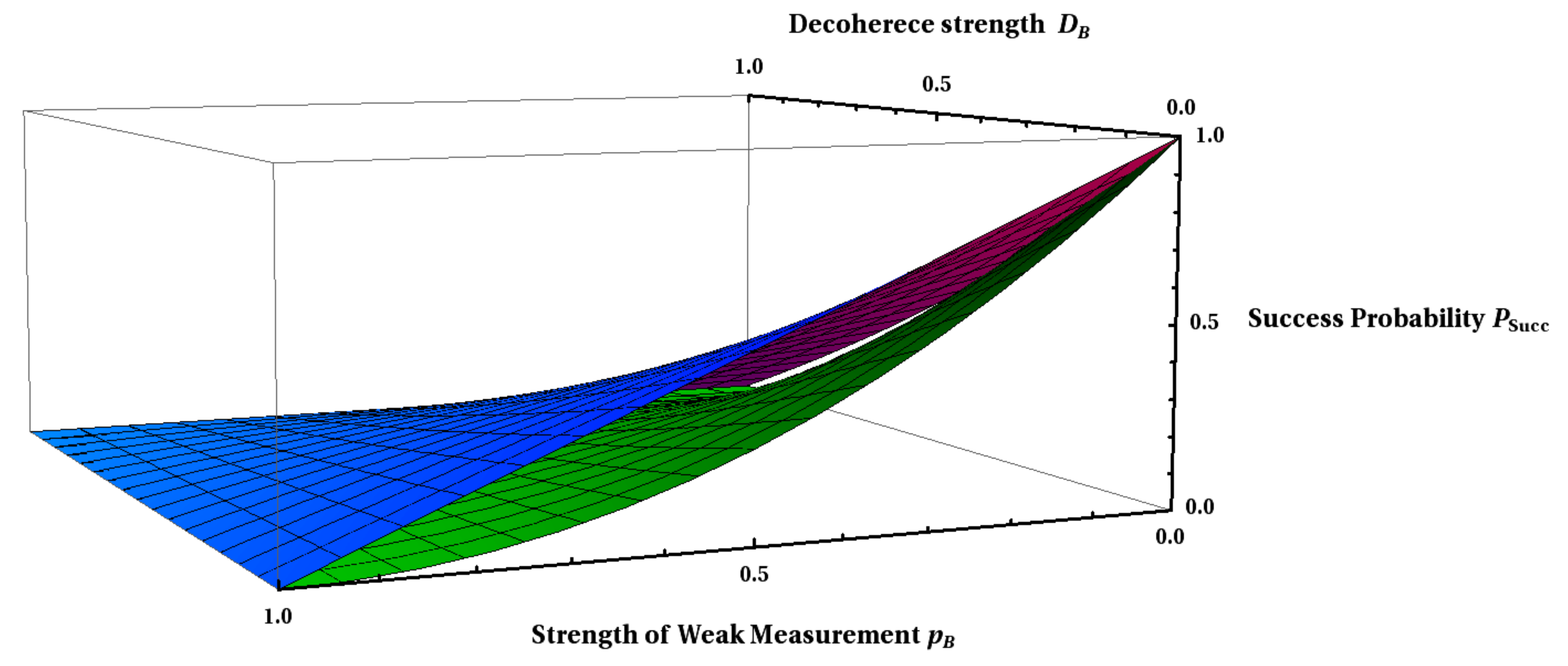}}}}

\caption{\footnotesize The upper surface represents the success probability $P_{Succ}^1$ (given by Eq.(\ref{PSucc1})) of {\it Case I}. The lower surface represents the success probability 
$P^{2}_{Succ}$ (given by Eq.(\ref{PSucc21})) of {\it Case II}.}
\end{figure}

Let us now consider the situation when  Alice prepares the two-qubit 
state in the class given by Eq.(\ref{anti-Parallel}). Without applying the 
weak measurement technique and in the presence of interaction of the 
environment with both  the particles, the FEF $\overline{f}_2$ is given 
by \cite{Som}
$ 1-D ~~~ \forall D\leq \frac{2}{3}$, and by
$ \frac{D}{2} ~~~~~~~~ \forall D\geq \frac{2}{3}$,
where we consider $D_1=D_2=D$. Comparing with the situation of {\it Case I}, when
only one of the qubits undergoes damping, one sees that $\overline{F}_2 \leq \overline{F}_1 ~~~~ \forall ~~ 2\sqrt{2}-2\leq D \leq \frac{8}{9}$, and
$\overline{F}_2 > \overline{F}1 ~~~~ \forall ~~ \frac{8}{9}< D\leq 1$,
with  $\overline{F}_2 =\frac{2 \overline{f}_2+1}{3}$, and 
 the FEF $\overline{f}_1$ is given by 
Eq.(\ref{FEFD1})).  $\overline{f}_1$ belongs to the classical region, i.e., 
$\overline{f}_1\leq 1/2$ for the choice of decoherence strength in the range 
given by $2\sqrt{2}-2\leq D\leq 1$.
When two qubits are allowed to interact with the environment, the fidelity 
$\overline{F}_2$ increases, but it never goes to non-classical region for the 
choice of $D$ from the region given by $2\sqrt{2}-2\leq D_2=D\leq 1$. 
However, when  weak measurement is applied with equal strength on both the 
qubits
for the state prepared in the class given by Eq.(\ref{anti-Parallel}), 
it remains unaffected. 
In order to see this, consider the weak measurement operation described
by Eq.(\ref{WKM0}). When a weak measurement is performed on a single qubit, say, the 2nd qubit  the states given by Eqs. (\ref{Parallel}) and (\ref{anti-Parallel}) become
$|\psi_{\pm}^M\rangle^W = M_{2,0}|\psi_{\pm}^M\rangle=\frac{|00\rangle_{12} \pm \sqrt{1-p_2} |11\rangle_{12}}{\sqrt{2-p_2}}$,
and 
$|\phi_\pm^M\rangle^W =M_{2,0}|\phi_{\pm}^M\rangle= \frac{|01\rangle_{12} \pm \sqrt{1-p_2} |10\rangle_{12}}{\sqrt{2-p_2}}$,
respectively. Note that the states $|\psi_{\pm}^M\rangle^W$ and $|\phi_{\pm}^M\rangle^W$ are connected by the unitary rotation $I\otimes\sigma_x$. Consequently, 
our protocol in reducing the effect of decoherence works for both the cases,
 and the success probability  is given by $\frac{(2-p_2)}{2}$. Now, if 
 weak measurements are performed on both qubits with equal strength $p_2$ (i.e., $p_1=p_2$), respectively,  the states given by Eqs. (\ref{Parallel}) and (\ref{anti-Parallel}) in this case become
$|\psi_{\pm}^M\rangle^{WW} = M_{1,0}|\psi_{\pm}^M\rangle^W=\frac{|00\rangle_{12} \pm (1-p_2) |11\rangle_{12}}{\sqrt{1+(1-p_2)^2}}$, and 
$|\phi_\pm^M\rangle^{WW} =M_{1,0}|\phi_{\pm}^M\rangle^W= \frac{|01\rangle_{12} \pm |10\rangle_{12}}{\sqrt{2}}$, respectively. 
In this case they are not unitarily related, and in fact, the state
$|\phi_\pm^M\rangle$ remains unaffected. 
Hence, the weak measurement technique is not useful for
 increasing the fidelity beyond the classical region for the state in the 
class given by Eq.(\ref{anti-Parallel}).

\section{Teleportation using Werner states}

The Werner state is given by
\begin{eqnarray}
\rho_W= \gamma ~ \eta_{\psi_\pm} + \frac{1-\gamma}{4} I,
\label{WSF1}
\end{eqnarray}
where $\eta_{\psi_\pm} = |\psi_\pm^M\rangle\langle \psi_\pm^M|$, where $|\psi_\pm^M\rangle$ is given by Eq.(\ref{Parallel}) and $\gamma$ is the mixing parameter lying between $0$ and $1$.  In the 
following analysis  we first discuss the effect on the teleportation fidelity 
due to 
the decoherence acting on  both qubits. We then discuss the case  when the 
technique of weak measurement is applied.

When both qubits interact with the environment via ADC, the FEF of the 
affected state $\rho_W^{DD}$ is given by
\begin{eqnarray}
\overline{f}_2(\rho_W^{DD}) &=& \frac{1}{4} \Big(D_2^2 (\gamma+1)-2 D_2 \gamma-2 (D_2-1)\gamma\nonumber \\
&& +\gamma+1 \Big),
\label{WSF_FEF_12_M}
\end{eqnarray}
where $D_1=D_2$, i.e., both qubits interact with the same environment. The 
corresponding fidelity, $\overline{F}_2(\rho_W^{DD})$ (=$\frac{2 \overline{f}_2(\rho_W^{DD}) + 1}{3}$) lies above the classical region for $0\leq D_2 \leq \frac{3\gamma -1}{1+\gamma} $ and $\gamma > \frac{1}{3}$.

\begin{figure}[h]
{\rotatebox{0}{\resizebox{9.0cm}{5.0cm}{\includegraphics{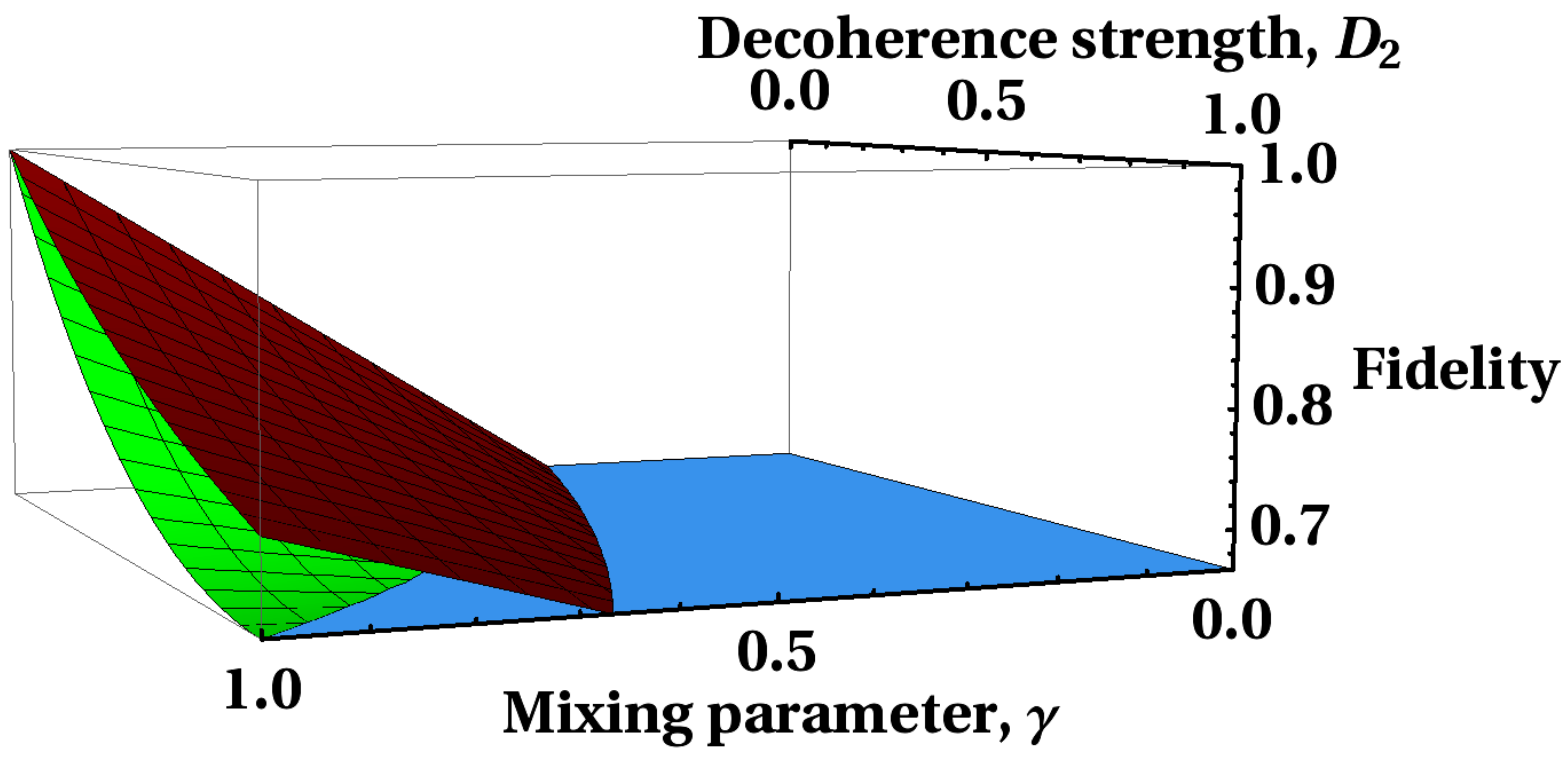}}}}

\caption{\footnotesize The upper surface represents the teleportation fidelity
when the technique of weak measurement is applied.
The lower surface represents the fidelity when weak measurement is not 
applied.}
\label{Fig_Werner}
\end{figure}

For the prepared state given by Eq.(\ref{WSF1}), when both qubits are affected by the environment Alice makes a weak measurement on each qubit and then 
sends the 2nd qubit to Bob. After getting the particle, Alice and Bob both apply reverse weak measurement on their respective particles. At the end they share 
the state $\rho_M^{RR}$ which has FEF given by
\begin{eqnarray}
f(\rho_M^{RR})&=&\big((q_2-1)^2 (D_2^2 (p_2-1)^2 (\gamma +1)\nonumber \\
&& +2 D_2 (p2-1) (\gamma -1)+\gamma +1)\nonumber \\
&&-4 (D_2-1) (p_2-1) (q_2-1) \gamma \nonumber \\
&&+(D_2-1)^2 (p_2-1)^2 (\gamma +1)\big)/\big(2 (q_2^2 (D_2^2 (p_2-1)^2 \nonumber \\
&& (\gamma +1)+2 D_2 (p_2-1) (\gamma -1)+\gamma +1) \nonumber \\
&&+q_2 (-2 D_2 (p_2-1) (p_2 \gamma +p_2-2)\nonumber \\
&&-2 p_2 (\gamma -1)-4)+p_2^2 (\gamma +1)-4 p_2+4)\big),
\end{eqnarray}
where $D_1=D_2$, $p_1=p_2$ and $q_1=q_2$. To get the optimal fidelity $F^O(\rho_M^{RR})$, we maximize $f(\rho_M^{RR})$ over the strength of the reverse weak 
measurement $q_2$. The optimal strength is given by
\begin{eqnarray}
q_2^O &=& \Big(D_2^2 (p_2-1)^2 (\gamma +1) -\big( (D_2-1)^2 (p_2-1)^2 (\gamma +1) \nonumber \\
&& (D_2^2 (p_2-1)^2 (\gamma +1)  +2 D_2 (p_2-1)    (\gamma -1)+\gamma +1)\big)^{1/2}\nonumber \\
&& +2 D_2 (p_2-1) (\gamma -1)+\gamma +1 \Big)/\Big(   D_2^2 (p_2-1)^2 (\gamma +1) \nonumber \\
&& +2 D_2 (p_2-1) (\gamma    -1)+\gamma +1 \Big).
\end{eqnarray}
The fidelity $F^O(\rho_M^{RR})$ lies above the classical region if $\gamma > \frac{1+D_2-D_2 p_2}{3-D_2+D_2p_2}$ ($D_2 < \frac{1-3\gamma}{(1+\gamma) (p_2-1)}$). The corresponding success probability is given by
\begin{eqnarray}
P_{Succ} &=& \frac{1}{4} ((q_2^O)^2 (D_2^2 (p_2-1)^2 (\gamma +1)+2 D_2 (p_2-1) (\gamma -1)\nonumber \\
&& +\gamma +1)+q_2^O (-2 D_2  (p_2-1) (p_2 \gamma +p_2-2)\nonumber \\
&&+p_2 (2-2 \gamma )-4) +p_2^2 (\gamma +1)-4 p_2+4).
\end{eqnarray}
When $D_2 < \frac{1-3\gamma}{(1+\gamma) (p_2-1)}$, the fidelity $F^O(\rho_M^{RR}) > \overline{F}_2(\eta_M^{DD})$, the weak measurement protocol is able
to enhance the teleportation fidelity. In Fig.(\ref{Fig_Werner}) we plot the
fidelity $F^O(\rho_M^{RR})$ versus the Werner state parameter $\gamma$ and the
decoherence strength $D_2$ for a particular value $p_2=0.4$ of the strength
of the weak measurement.  Note that even for large values of the mixing
parameter $\gamma$ and the magnitude of decoherence $D_2$, the fidelity
in the quantum region is achieved through the technique of weak measurement. 

 \section{Conclusions}
 
To summarize, in the present work, we propose a method for maintaining teleportation
fidelity over the classical region through noisy channels using the technique of weak
measurements. We reduce the loss of information about the unknown state due to 
interaction  with the environment via amplitude damping channel with the help 
of weak measurement and reversal of weak measurement. 
We find the optimal 
strength of reversing the measurement for which the loss is minimum. 
Our results are exemplified by two classes of two-qubit states, {\it viz.}
maximally entangled pure states, as well as
the mixed Werner state.
For the 
prepared two-qubit states given by Eqs.(\ref{Parallel}) and 
(\ref{anti-Parallel}), we show that when only one particle (say, the $2$nd 
qubit which Alice sends to  Bob over the environment) interacts with 
environment, the weak measurement technique is able to enhance the 
teleportation fidelity arbitrarily close to $1$. This result holds good
for all maximally entangled states, as well as for all values of the
decoherence parameter. In this case even without 
performing the weak measurement before sending the $2$nd particle, i.e., 
$p_2=0$, Bob is able to enhance the fidelity to the  quantum region by making 
a  weak measurement with strength given by Eq.(\ref{Optimal1}). 
However, without applying the weak measurement technique by Alice and Bob, the 
teleportation fidelity lies in the classical region for the choice of 
decoherence strength chosen from the region $2\sqrt{2}-2\leq D_2\leq 1$. 

Next, 
when the environment effects both the particles, the weak measurement 
technique protects the information  for the initially prepared state given 
by Eq.(\ref{Parallel}), but fails to do so for the state given by Eq.(\ref{anti-Parallel}). Note that though the states  Eq.(\ref{Parallel}) and 
Eq.(\ref{anti-Parallel}) are unitarily equivalent to each other,  the nature of post-selection employed by us in the process of the weak measurement 
is unable to impact the state in the latter case due to its chosen structure.  
We also show that by increasing the strength of weak measurement, 
the success probability (which arises as a consequence of the cancellation of 
the protocol when the detector clicks) decreases. The success of enhancing 
teleportation fidelity is larger when one qubit interacts compared to the
case when both qubits interact with the environment. We also employ our
protocol for the Werner state and show that for a large range of the mixing parameter as well as the decoherence strength, the technique of weak measurement
is able to improve the teleportation fidelity beyond the classical region.

{\it  Acknowledgements:}  The authors thank Satyabrata Adhikari and Guruprasad
Kar for helpful discussions. TP thanks UGC, India for financial support. ASM acknowledges support from the DST project no. SR/S2/PU-16/2007.

\end{document}